\documentstyle[twoside,epsf,psfig,amssymb]{article}

\catcode`\@=11
\long\def\@makefntext#1{
\protect\noindent \hbox to 3.2pt {\hskip-.9pt  
$^{{\eightrm\@thefnmark}}$\hfil}#1\hfill}		

\def\@makefnmark{\hbox to 0pt{$^{\@thefnmark}$\hss}}	
	
\def\ps@myheadings{\let\@mkboth\@gobbletwo
\def\@oddhead{\hbox{}
\rightmark\hfil\eightrm\thepage}   
\def\@oddfoot{}\def\@evenhead{\eightrm\thepage\hfil
\leftmark\hbox{}}\def\@evenfoot{}
\def\sectionmark##1{}\def\subsectionmark##1{}}

\oddsidemargin=\evensidemargin
\newcounter{sectionc}\newcounter{subsectionc}\newcounter{subsubsectionc}
\renewcommand{\section}[1] {\vspace{12pt}\addtocounter{sectionc}{1} 
\setcounter{subsectionc}{0}\setcounter{subsubsectionc}{0}\noindent 
	{\tenbf\thesectionc. #1}\par\vspace{5pt}}
\renewcommand{\subsection}[1] {\vspace{12pt}\addtocounter{subsectionc}{1} 
\setcounter{subsubsectionc}{0}\noindent 
{\bf\thesectionc.\thesubsectionc. {\kern1pt \bfit #1}}\par\vspace{5pt}}
\renewcommand{\subsubsection}[1] {\vspace{12pt}\addtocounter{subsubsectionc}{1}
	\noindent{\tenrm\thesectionc.\thesubsectionc.\thesubsubsectionc.
	{\kern1pt \tenit #1}}\par\vspace{5pt}}
\newcommand{\nonumsection}[1] {\vspace{12pt}\noindent{\tenbf #1}
	\par\vspace{5pt}}

\newcounter{appendixc}
\newcounter{subappendixc}[appendixc]
\newcounter{subsubappendixc}[subappendixc]
\renewcommand{\thesubappendixc}{\Alph{appendixc}.\arabic{subappendixc}}
\renewcommand{\thesubsubappendixc}
	{\Alph{appendixc}.\arabic{subappendixc}.\arabic{subsubappendixc}}

\renewcommand{\appendix}[1] {\vspace{12pt}
        \refstepcounter{appendixc}
        \setcounter{figure}{0}
        \setcounter{table}{0}
        \setcounter{lemma}{0}
        \setcounter{theorem}{0}
        \setcounter{corollary}{0}
        \setcounter{definition}{0}
        \setcounter{equation}{0}
        \renewcommand{\thefigure}{\Alph{appendixc}.\arabic{figure}}
        \renewcommand{\thetable}{\Alph{appendixc}.\arabic{table}}
        \renewcommand{\theappendixc}{\Alph{appendixc}}
        \renewcommand{\thelemma}{\Alph{appendixc}.\arabic{lemma}}
        \renewcommand{\thetheorem}{\Alph{appendixc}.\arabic{theorem}}
        \renewcommand{\thedefinition}{\Alph{appendixc}.\arabic{definition}}
        \renewcommand{\thecorollary}{\Alph{appendixc}.\arabic{corollary}}
        \renewcommand{\theequation}{\Alph{appendixc}.\arabic{equation}}
        \noindent{\tenbf Appendix \theappendixc #1}\par\vspace{5pt}}
\newcommand{\subappendix}[1] {\vspace{12pt}
        \refstepcounter{subappendixc}
        \noindent{\bf Appendix \thesubappendixc. {\kern1pt \bfit #1}}
	\par\vspace{5pt}}
\newcommand{\subsubappendix}[1] {\vspace{12pt}
        \refstepcounter{subsubappendixc}
        \noindent{\rm Appendix \thesubsubappendixc. {\kern1pt \tenit #1}}
	\par\vspace{5pt}}

\newcommand{\textlineskip}{\baselineskip=13pt}
\newcommand{\smalllineskip}{\baselineskip=10pt}

\newcommand{\copyrightheading}[1]
	{\vspace*{-2.5cm}\smalllineskip{\flushleft
	{\footnotesize International Journal of Modern Physics D, #1}\\
	{\footnotesize \copyright\kern2pt World Scientific Publishing
	 Company}\\
	 }}

\newcommand{\publisher}[2]{{\begin{center}\footnotesize\smalllineskip 
	Received #1\\
	Revised #2
	\end{center}
	}}

\def\abstracts#1#2#3{{
	\centering{\begin{minipage}{4.5in}\footnotesize\baselineskip=10pt
	\parindent=0pt #1\par 
	\parindent=15pt #2\par
	\parindent=15pt #3
	\end{minipage}}\par}} 

\def\keywords#1{{
	\centering{\begin{minipage}{4.5in}\footnotesize\baselineskip=10pt
	{\footnotesize\it Keywords}\/: #1
	 \end{minipage}}\par}}

\renewenvironment{thebibliography}[1]
        {\frenchspacing
	 \ninerm\baselineskip=11pt
         \begin{list}{\arabic{enumi}.}
        {\usecounter{enumi}\setlength{\parsep}{0pt}     
	 \setlength{\leftmargin 12.7pt}{\rightmargin 0pt}
         \setlength{\itemsep}{0pt} \settowidth
	{\labelwidth}{#1.}\sloppy}}{\end{list}}

\newcounter{itemlistc}
\newcounter{romanlistc}
\newcounter{alphlistc}
\newcounter{arabiclistc}

\newcommand{\fcaption}[1]{
        \refstepcounter{figure}
        \setbox\@tempboxa = \hbox{\footnotesize Fig.~\thefigure. #1}
        \ifdim \wd\@tempboxa > 5in
           {\begin{center}
        \parbox{5in}{\footnotesize\smalllineskip Fig.~\thefigure. #1}
            \end{center}}
        \else
             {\begin{center}
             {\footnotesize Fig.~\thefigure. #1}
              \end{center}}
        \fi}

\newcommand{\tcaption}[1]{
        \refstepcounter{table}
        \setbox\@tempboxa = \hbox{\footnotesize Table~\thetable. #1}
        \ifdim \wd\@tempboxa > 5in
           {\begin{center}
        \parbox{5in}{\footnotesize\smalllineskip Table~\thetable. #1}
            \end{center}}
        \else
             {\begin{center}
             {\footnotesize Table~\thetable. #1}
              \end{center}}
        \fi}

\def\@citex[#1]#2{\if@filesw\immediate\write\@auxout
	{\string\citation{#2}}\fi
\def\@citea{}\@cite{\@for\@citeb:=#2\do
	{\@citea\def\@citea{,}\@ifundefined
	{b@\@citeb}{{\bf ?}\@warning
	{Citation `\@citeb' on page \thepage \space undefined}}
	{\csname b@\@citeb\endcsname}}}{#1}}

\newif\if@cghi
\def\cite{\@cghitrue\@ifnextchar [{\@tempswatrue
	\@citex}{\@tempswafalse\@citex[]}}
\def\citelow{\@cghifalse\@ifnextchar [{\@tempswatrue
	\@citex}{\@tempswafalse\@citex[]}}
\def\@cite#1#2{{$\null^{#1}$\if@tempswa\typeout
	{IJCGA warning: optional citation argument 
	ignored: `#2'} \fi}}

\def\pmb#1{\setbox0=\hbox{#1}
	\kern-.025em\copy0\kern-\wd0
	\kern.05em\copy0\kern-\wd0
	\kern-.025em\raise.0433em\box0}

\def\fnt#1#2{\footnotetext{\kern-.3em
	{$^{\mbox{\scriptsize #1}}$}{#2}}}

\def\fpage#1{\begingroup
\voffset=.3in
\thispagestyle{empty}\begin{table}[b]\centerline{\footnotesize #1}
	\end{table}\endgroup}

\def\runninghead#1#2{\pagestyle{myheadings}
\markboth{{\protect\footnotesize\it{\quad #1}}\hfill}
{\hfill{\protect\footnotesize\it{#2\quad}}}}
\font\tenrm=cmr10
\font\tenit=cmti10 
\font\tenbf=cmbx10
\font\bfit=cmbxti10 at 10pt
\font\ninerm=cmr9

\font\eightrm=cmr8

\def\qed{\hbox{${\vcenter{\vbox{	          
   \hrule height 0.4pt\hbox{\vrule width 0.4pt height 6pt
   \kern5pt\vrule width 0.4pt}\hrule height 0.4pt}}}$}}

\begin{document}
\setlength{\textheight}{7.7truein}    

\runninghead{J. A. Belinch\'on.} {Wheeler-DeWitt Equation with Variable Constants.}

\normalsize\textlineskip
\thispagestyle{empty}
\setcounter{page}{1}

\copyrightheading{Vol.~11, No.~4 (2002) 527--543}	

\vspace*{0.88truein}

\fpage{527}

\title{Wheeler-DeWitt equation with variable constants.}
\centerline{\bf WHEELER-DEWITT EQUATION WITH VARIABLE CONSTANTS.}
\vspace*{0.37truein}

\centerline{\footnotesize JOS\'E ANTONIO BELINCH\'ON}
\vspace*{0.015truein}

\centerline{\footnotesize\it Grupo Inter-Universitario de An\'alisis Dimensional}
\centerline{\footnotesize\it Dept. F\'{\i}sica ETS Arquitectura}
\centerline{\footnotesize\it UPM Av. Juan de Herrera N 4 Madrid 28040 Espa\~{n}a}
\centerline{\footnotesize\it E-mail: jabelinchon@members.nyas.org}
\vspace*{10pt}
\vspace*{0.225truein}
\publisher{(1 June 2001)}{(24 September 2001)}
\vspace*{0.21truein}

\abstracts{In this paper we study how all the physical ``constants" vary in the framework described by a model in which we have taken into account the generalize conservation principle for its stress-energy tensor. This possibility enable us to take into account the adiabatic matter creation in order to get rid of the entropy problem. We try to generalize this situation by contemplating multi-fluid components. To validate all the obtained results we explore the possibility of considering the variation of  the``constants" in the quantum cosmological scenario described by the Wheeler-DeWitt equation. For this purpose we explore the Wheeler-DeWitt equation in different contexts but from
a dimensional point of view. We end by presenting the Wheeler-DeWitt equation
in the case of considering all the constants varying. The quantum potential is
obtained and the tunneling probability is studied.}{}{}
\vspace*{10pt}

\setcounter{page}{527}

\section{Introduction.}

In a recent paper \cite{T1} we studied the behaviour of the ``constants''
$G,c,\Lambda$ and $\hbar$ in the framework described by a cosmological model
with FRW symmetries in which we have imposed the equation of state
$\rho=a\theta^{4}$ for the energy density. In this way a possible variation of
the ``constant'' $\hbar$ is reflected in the field equations. Formally, the
 obtained result only is valid for a determinate kind of matter, radiation
predominance, but we extrapolated the result for other kind of matter, dust,
strings, ultrastiff matter etc... checking that such extrapolation is right
since any equations of physics as Maxwell, Schr\"{o}dinger or Klein-Gordon
equations remain gauge invariant if we introduce into them the ``constants''
varying for any kind of matter considered in the field equations.

Now we would like to validate the obtained result and to find the behaviour
for $\hbar$ as well as for the rest of the ``constants'' through a equation in
which ``functions'' (``constants'') appear independently of the imposed
equation of state. For this purpose we resort to the Wheeler-DeWitt (W-DW)
equation in a suitable minisuperspace, since in this equation 
such ``constants'' for any kind of matter always appear. Therefore, in this paper we try to
study the behaviour of these ``constants'' in the framework described by
quantum cosmology.

The paper is organized as follows: In section $2$ we review briefly the
obtained result
 in reference \cite{T1} about the variation of the constants and we
shall introduce new results that will show us how the constants vary when we
consider mechanism of adiabatic matter creation. We shall explore the
possibility of considering the variation of constants in the multi-fluid
scenario, i.e. the stress-energy tensor is defined by various kinds of matter
as for example radiation, dust and strings. Once these results have been
exposed, in section $3$ we shall study the Wheeler-DeWitt equation in the
minisuperspace approximation i.e. we shall impose FRW symmetries. Obviously,
all the ``constants'' will be taken into account when we write such equation
and we shall study it from the dimensional point of view. We will obtain the
reduced form of the W-DW equation i.e. the Schr\"{o}dinger equation, in such a
way that all the quantum procedure could be used. We shall check that the
 obtained results in section $2$ remain gauge invariant the W-DW equation. In
section $4$ we will present another model described by a scalar field, also
studying it from a dimensional point of view. Finally, in section $5$ we shall
study both equation, W-DW equation as well as the Schr\"{o}dinger's one (its
reduced form) with all the constants varying. We will obtain its potential and
the tunneling probability is studied. To calculate such probability we employ
the Gamow's classical formula in which all the ``constants'' appear, not only
the potential as in previous proposals, since we believe that their variations
must be reflected in such calculation.

\section{\textbf{Variable constants.}}

In a recent paper\cite{T1} we studied a flat cosmological model with FRW
symmetries where the energy-momentum tensor is defined by a perfect fluid. As
we have indicated in the introduction, in order to reflect the possible variation of
``constant'' $\hbar$ into the field equations, we impose the equation of state
$\rho=a\theta^{4}$ for the energy density $\rho$, where $\theta$ stands for
the temperature and $a$ is the radiation constant.

The field equations are:
\begin{equation}
2\frac{f\,^{\prime\prime}}{f\,}+\frac{(f\,^{\prime})^{2}}{f\,^{2}}
=-\frac{8\pi G(t)}{c(t)^{2}}p+c(t)^{2}\Lambda(t),\label{p1}%
\end{equation}%
\begin{equation}
3\frac{(f\,^{\prime})^{2}}{f\,^{2}} =\frac{8\pi G(t)}{\,c(t)^{2}}\rho
+c(t)^{2}\Lambda(t).\label{p2}%
\end{equation}

The differential equation that describes the variation of all the
``constants'' is obtained from the condition
\begin{equation}
div\left(  \frac{8\pi G}{c^{4}}T_{i}^{j}+\delta_{i}^{j}\Lambda\right)  =0,
\end{equation}
that simplified is:
\begin{equation}
T_{i;j}^{j}-\left(  \frac{4c_{,j}}{c}-\frac{G_{,j}}{G}\right)  T_{i}^{j}%
+\frac{c^{4}(t)\delta_{i}^{j}\Lambda_{,j}}{8\pi G}=0
\end{equation}
or equivalently:
\begin{equation}
4\frac{\theta^{\prime}}{\theta}-3\left[  \frac{c^{\prime}}{c}+\frac
{\hbar^{\prime}}{\hbar}\right]  +3(\omega+1)H+\frac{15\Lambda^{\prime}%
c^{7}\hbar^{3}}{8\pi^{3}Gk_{B}^{4}\theta^{4}}+\frac{G^{\prime}}{G}%
-4\frac{c^{\prime}}{c}=0.\label{BER4}%
\end{equation}

To solve this equation we imposed two simplifying hypotheses, the first one,
that the relation $G/c^{2}=B$ (where $B$ is a $const.$) remains constant, and
the second one, that the cosmological ``constant'' verifies the relation
$\Lambda\propto\frac{d}{c^{2}t^{2}}$ with $d\in\mathbb{R},$ while $div\left(
T_{i}^{j}\right)  \neq0$, in this way the equation was solved perfectly. Since
the constant $B$ has dimensions $\left[  B\right]  =LM^{-1}$ we can get the
dimensionless monomia $\pi_{1}=\frac{\rho Bt^{2}}{b}$ where $b\in\mathbb{R}.$
With these hypotheses and $\pi_{1}$ equation (\ref{BER4}) simplifies to:
\begin{equation}
4\frac{\theta^{\prime}}{\theta}-3\left[  \frac{c^{\prime}}{c}+\frac
{\hbar^{\prime}}{\hbar}\right]  +3(\omega+1)H-\frac{15d}{4\pi^{3}}\frac
{c^{4}\left[  c^{\prime}t+c\right]  \hbar^{3}}{Gk_{B}^{4}\theta^{4}t^{3}%
}+\frac{G^{\prime}}{G}-4\frac{c^{\prime}}{c}=0,\label{p5}%
\end{equation}
that has no immediate integration. We have to take into account the field
equations (\ref{p2})
\begin{equation}
3H^{2}=8\pi bt^{-2}+dt^{-2},\label{LUC1}%
\end{equation}
from this we get $f=K_{\varkappa}t^{\varkappa}$ where $\varkappa=\left(
\frac{8\pi b+d}{3}\right)  ^{\frac{1}{2}}$ and substituting in (\ref{p5})
together with $G=Bc^{2}$ we get
\begin{equation}
4\frac{\theta^{\prime}}{\theta}-3\left[  \frac{c^{\prime}}{c}+\frac
{\hbar^{\prime}}{\hbar}\right]  +\frac{3(\omega+1)\varkappa}{t}-\frac
{15d}{4\pi^{3}}\frac{c^{2}\left[  c^{\prime}t+c\right]  \hbar^{3}}{Bk_{B}%
^{4}\theta^{4}t^{3}}-2\frac{c^{\prime}}{c}=0,\label{LUCIL1}%
\end{equation}
i.e. one equation with $3$ unknowns.

If we want to integrate equation (\ref{LUCIL1}) we have to take a decision on
the behavior of the constant $\hbar.$

Taking $c\hbar=const.\approx\hbar=\frac{A}{c}$ then $\frac{\hbar^{\prime}%
}{\hbar}=-\frac{c^{\prime}}{c}$ yielding:
\begin{equation}
4\frac{\theta^{\prime}}{\theta}+\frac{3(\omega+1)\varkappa}{t}-\frac{15d}%
{4\pi^{3}}\frac{A^{3}\left[  c^{\prime}t+c\right]  }{cBk_{B}^{4}\theta
^{4}t^{3}}-2\frac{c^{\prime}}{c}=0.
\end{equation}
Also if $\rho=a\theta^{4}$ and $\rho=\frac{b}{Bt^{2}}\Longrightarrow\frac
{b}{Bt^{2}}=a\theta^{4}$ we have:
\begin{equation}
k_{B}\theta=\left(  \frac{15c^{3}(t)\hbar^{3}(t)b}{\pi^{2}Bt^{2}}\right)
^{\frac{1}{4}}=\left(  \frac{15A^{3}b}{\pi^{2}B}\right)  ^{\frac{1}{4}%
}t^{-1/2}.
\end{equation}
And substituting into the previous equation we get :
\begin{equation}
\frac{-2}{t}+\frac{3(\omega+1)\varkappa}{t}-\frac{15d}{4\pi^{3}}\frac
{A^{3}\left[  c^{\prime}t+c\right]  }{cBt^{3}\frac{15A^{3}b}{\pi^{2}Bt^{2}}%
}-2\frac{c^{\prime}}{c}=0,
\end{equation}
and simplifying
\begin{equation}
\frac{-2}{t}+\frac{3(\omega+1)\varkappa}{t}-\frac{d}{4\pi b}\left[
\frac{c^{\prime}}{c}+\frac{1}{t}\right]  -2\frac{c^{\prime}}{c}=0,\label{REL2}%
\end{equation}
therefore we get a very simple differential equation:
\begin{equation}
\frac{c^{\prime}}{c}=\left[  \frac{12\pi b(\omega+1)\varkappa-8\pi b-d}{8\pi
b+d}\right]  \frac{1}{t}%
\end{equation}
integrating it we obtain easily:
\begin{equation}
c=K_{\xi}t^{\xi}%
\end{equation}
where $\xi=\left[  \frac{12\pi b(\omega+1)\varkappa-8\pi b-d}{8\pi
b+d}\right]  $.

We can consider another possibility. Take the group of governing quantities
$\frak{M=M}$$\left\{  K_{\varkappa},A,t\right\}  $ where $K_{\varkappa}$ is
the proportionality constant obtained from $f=K_{\varkappa}t^{\varkappa}$ and
$A$ is the constant establishing the relation between $\hbar$ and $c$. The
results obtained by means of the gauge relations are:
\begin{equation}%
\begin{array}
[c]{l}%
G\propto K_{\varkappa}^{6}A^{-1}t^{6\varkappa-4}\\
c\propto K_{\varkappa}t^{\varkappa-1}\\
\hbar\propto K_{\varkappa}^{-1}At^{1-\varkappa}\\
k_{B}\theta\propto K_{\varkappa}^{-1}At^{-\varkappa}\\
\rho\propto K_{\varkappa}^{-4}At^{-4\varkappa}\\
m_{i}\propto K_{\varkappa}^{3}At^{-3\varkappa+2}\\
\Lambda\propto K_{\varkappa}^{-2}t^{-2\varkappa}\\
e^{2}\varepsilon_{0}^{-1}\propto A
\end{array}
\label{OLG3}%
\end{equation}
where $m_{i}$ comes from the energy density definition $\rho_{E}=\frac
{nm_{i}c^{2}}{f^{3}}$ ($n$ stands for the particles number) and $e^{2}%
\varepsilon_{0}^{-1}$ from the definition of the fine structure constant
$\alpha$. We can check that we recover the general covariance property
$\frac{G}{c^{2}}=const.$ \ iif $\varkappa=\frac{1}{2}$. Similarly we can see
that the following relations are satisfied: $\rho=a\theta^{4},$ $\rho=Af^{-4}$
(equivalent to $div(T)=0),$ $\Lambda\propto f^{-2}$ and $f=ct$ (no horizon
problem). And finally $e^{2}\varepsilon_{0}^{-1}\propto const$. In this way
the fine structure constant $\alpha\propto\frac{e^{2}}{\varepsilon_{0}c\hbar
}=const.,$ is a true constant. With the value $\varkappa=\frac{1}{2}$ we get
\begin{equation}%
\begin{array}
[c]{l}%
c\propto t^{-1/2},\qquad\hbar\propto t^{1/2},\qquad G\propto t^{-1},\qquad
k_{B}\theta\propto t^{-1/2}\\
\qquad f\propto t^{1/2},\qquad\rho\propto t^{-2},\qquad m_{i}\propto t^{1/2}%
\end{array}
\label{OLG4}%
\end{equation}

Another alternative is to consider the governing quantities $\frak{M=M}$
$\left\{  B,A,t\right\}  $ such that $\left[  A\right]  =\left[  A_{\omega
}\right]  $ with $\omega=\frac{1}{3}$. So that we get the same problem as with
the condition $div(T)=0$ (Note $\rho=A_{\omega}f^{-3(\omega+1)}$) This would
imply no previous hypothesis on the behavior of $\hbar$. With these results we
have seen that the equations of the physic like Maxwell, Schr\"{o}dinger or
Klein-Gordon equations remain gauge invariant. We emphasize that in such work
we determined the behaviour of $\hbar$ without any doubt within this framework
(see\cite{T1} for details)$.$

In this paper we would like to study briefly the important case in which the
condition $div\left(  T_{i}^{j}\right)  =0$ is taken into account in such a
way that adiabatic matter creation can be taken into account, in order to get
rid of the entropy problem since in\cite{T1} this problem was no solved.

With these new assumptions the field equations that now govern the model are
as follows:
\begin{equation}
2\frac{f\,^{\prime\prime}}{f\,}+\frac{(f\,^{\prime})^{2}}{f\,^{2}}=-\frac{8\pi
G(t)}{c^{2}(t)}(p+p_{c})+c^{2}(t)\Lambda(t),\label{e1}%
\end{equation}%
\begin{equation}
3\frac{(f\,^{\prime})^{2}}{f\,^{2}}=\frac{8\pi G(t)}{\,c^{2}(t)}\rho
+c^{2}(t)\Lambda(t),\label{e2}%
\end{equation}%
\begin{equation}
n^{\prime}+3nH=\psi,\label{e3}%
\end{equation}
and taking into account our general assumption i.e.
\begin{equation}
T_{i;j}^{j}-\left(  \frac{4c_{,j}}{c}-\frac{G_{,j}}{G}\right)  T_{i}^{j}%
+\frac{c^{4}(t)\delta_{i}^{j}\Lambda_{,j}}{8\pi G}=0
\end{equation}
with $T_{i;j}^{j}=0,$ this brings us to obtain two equations
\begin{equation}
\rho^{\prime}+3\left(  \rho+p+p_{c}\right)  H=0\label{maciza1}%
\end{equation}
and
\begin{equation}
\frac{\Lambda^{\prime}c^{4}}{8\pi G\rho}+\frac{G^{\prime}}{G}-4\frac
{c^{\prime}}{c}=0\label{maciza2}%
\end{equation}
where $n$ measures the particles number density, $\psi$ is the function that
measures the matter creation, $H=f^{\prime}/f$ represents the Hubble parameter
($f$ is the scale factor that appears in the metric), $p$ is the thermostatic
pressure, $\rho$ is energy density and $p_{c}$ is the pressure that generates
the matter creation.

The creation pressure $p_{c}$ depends on the function $\psi$. For adiabatic
matter creation this pressure takes the following form:
\begin{equation}
p_{c}=-\left[  \frac{\rho+p}{3nH}\psi\right]  .\label{w2}%
\end{equation}
The state equation that we next use is the well-known expression
\begin{equation}
p=\omega\rho\label{w3}%
\end{equation}
where $\omega=const.$ $\omega\in(-1,1]$. We assume that this function follows
the law:
\begin{equation}
\psi=3\beta nH,\label{w5}%
\end{equation}
(see \cite{T2}) where $\beta$ is a dimensionless constant (if $\beta=0$ then
there is no matter creation since $\psi=0)$. The generalized principle of
conservation $T_{i;j}^{j}=0,$ for the stress-energy tensor (\ref{maciza1})
brings us to:
\begin{equation}
\rho^{\prime}+3(\omega+1)\rho\frac{f^{\prime}}{f}=(\omega+1)\rho\frac{\psi}%
{n}.\label{w4}%
\end{equation}
By integrating equation (\ref{w4}) we obtain the following relation between
the energy density and the scale factor and which is more important, the
constant of integration that we shall need for our subsequent calculations:
\begin{equation}
\rho=A_{\omega,\beta}f^{-3(\omega+1)(1-\beta)},\label{e4}%
\end{equation}
where $A_{\omega,\beta}$ is the integration constant that depends on the
equation of state that we want to consider i.e. constant $\omega$ and constant
$\beta$ that controls the matter creation, $\left[  A_{\omega,\beta}\right]
=L^{3(\omega+1)(1-\beta)-1}MT^{-2}$. With this constant of integration and
taking into account the hypothesis about the relation $G/c^{2}=B,$ our purpose
is to show that no more hypothesis are necessary to solve the differential
equations that govern the model. Therefore the set of governing parameters are
now: $\frak{M=M}$$\left\{  A_{\omega,\beta},B,t\right\}  ,$ that brings us to
obtain the next relations:
\begin{equation}%
\begin{array}
[c]{l}%
G\propto A_{\omega,\beta}^{\frac{2}{\gamma+1}}B^{\frac{2}{\gamma+1}+1}%
t^{\frac{2(1-\gamma)}{\gamma+1}},\\
c\propto A_{\omega,\beta}^{\frac{1}{\gamma+1}}B^{\frac{1}{\gamma+1}}%
t^{\frac{(1-\gamma)}{\gamma+1}},\\
\hbar\propto A_{\omega,\beta}^{\frac{3}{\gamma+1}}B^{\frac{3}{\gamma+1}%
-1}t^{\frac{5-\gamma}{\gamma+1}},\\
m_{i}\propto A_{\omega,\beta}^{\frac{1}{\gamma+1}}B^{-\frac{\gamma}{\gamma+1}%
}t^{\frac{2}{\gamma+1}},\\
e^{2}\varepsilon_{0}^{-1}\propto A_{\omega,\beta}^{\frac{\gamma-3}{\gamma
+1}+1}B^{\frac{\gamma-3}{\gamma+1}}t^{2\frac{\gamma-3}{\gamma+1}},\\
\rho\propto B^{-1}t^{-2},\\
f\propto A_{\omega,\beta}^{\frac{1}{\gamma+1}}B^{\frac{1}{\gamma+1}}%
t^{\frac{2}{\gamma+1}},\\
k_{B}\theta\propto A_{\omega,\beta}^{\frac{3}{\gamma+1}}B^{\frac{3}{\gamma
+1}-1}t^{\frac{4-2\gamma}{\gamma+1}},\\
a^{-1/4}s\propto A_{\omega,\beta}^{\frac{3}{\gamma+1}}B^{\frac{3}{\gamma
+1}-\frac{3}{4}}t^{\frac{6}{\gamma+1}-\frac{3}{2}},\\
\Lambda\propto A_{\omega,\beta}^{\frac{-2}{\gamma+1}}B^{\frac{-2}{\gamma+1}%
}t^{\frac{-4}{\gamma+1}},\\
q=\frac{\gamma-1}{2}%
\end{array}
\label{Tabla1}%
\end{equation}
where $\gamma=3(\omega+1)(1-\beta)-1,$ and $q$ is the deceleration parameter.
If $\omega=1/3$ then it is observed that the entropy is not constant
\begin{equation}
a^{-1/4}s\propto t^{\frac{3}{2(1-\beta)}-\frac{3}{2}},
\end{equation}
we can check that the next results are verified: we see that $\frac{G}{c^{2}%
}=B$ $\forall\beta$ (trivially), $\rho=a\theta^{4}$ $\forall\beta$, $f=ct$
$\forall\beta$, $\Lambda\propto\frac{1}{c^{2}t^{2}}\propto f^{-2}$ while the
relation $\hbar c\neq const.$ since it depends on $\beta.$ We also can check that
our model has no the so called Planck's problem since the Planck system
behaves now as:
\begin{equation}%
\begin{array}
[c]{l}%
l_{p}=\left(  \frac{G\hbar}{c^{3}}\right)  ^{1/2}\approx f(t),\\
m_{p}=\left(  \frac{c\hbar}{G}\right)  ^{1/2}\approx f(t),\\
t_{p}=\left(  \frac{G\hbar}{c^{5}}\right)  ^{1/2}\approx t,
\end{array}
\end{equation}
since the radius of the Universe $f(t)$ at Planck's epoch coincides with the
Planck's length $f(t_{p})\approx l_{p}$, while the energy density at Planck's
epoch coincides with the Planck's energy density $\rho(t_{p})\approx\rho
_{p}\approx t^{-2},$ where $\rho_{p}=m_{p}c^{2}/l_{p}^{3}.$ See\cite{T2} for
more details and the followed method etc...

It is observed from (\ref{Tabla1}) that if we make $\beta=0$ the following set
of solutions are obtained:
\begin{equation}%
\begin{array}
[c]{|l|r|r|r|r|r|r|}%
\omega & 1 & 2/3 & 1/3 & 0 & -1/3 & -2/3\\\hline
f & 1/3 & 2/5 & 1/2 & 2/3 & 1 & 2\\\hline
\rho & -2 & -2 & -2 & -2 & -2 & -2\\\hline
\theta & -1 & -4/5 & -1/2 & 0 & 1 & 4\\\hline
s & -1/2 & -3/10 & 0 & 1/2 & 3/2 & 9/2\\\hline
G & -4/3 & -6/5 & -1 & -2/3 & 0 & 2\\\hline
c & -2/3 & -3/5 & -1/2 & -1/3 & 0 & 1\\\hline
\hbar & 0 & 1/5 & 1/2 & 1 & 2 & 5\\\hline
m_{i} & 1/3 & 2/5 & 1/2 & 2/3 & 1 & 2\\\hline
\Lambda & -2/3 & -4/5 & -1 & -4/3 & -2 & -4\\\hline
e^{2}\varepsilon_{0}^{-2} & -2/3 & -2/5 & 0 & 2/3 & 2 & 6\\\hline
q & 5/2 & 3/2 & 1 & 1/2 & 0 & -1/2
\end{array}
\label{Tabla2}%
\end{equation}
with: $(\omega=-1)$ corresponds to de Sitter (false vacuum) represented by the
cosmological constant (special case), $(\omega=-\frac{2}{3})$ for domain
walls, $(\omega=-\frac{1}{3})$ for strings, $(\omega=0)$ for dust (matter
predominance), $(\omega=\frac{1}{3})$ for radiation or ultrarelativistic gases
(radiation predominance), $(\omega=\frac{2}{3})$ for perfect gases,
$(\omega=1)$ for ultra-stiff matter. For example, if we take the case
$\omega=0$ the table (\ref{Tabla2}) tells us that
\[
f\propto t^{2/3},\qquad\rho\propto t^{-2},\qquad\theta\propto t^{0}%
=const.,.....
\]%
\[
G\propto t^{-2/3},\qquad c\propto t^{-1/3},\qquad\hbar\propto t,...etc.
\]

This table tells us too that if we want that our universe accelerates then we
have to impose that $-1\leq\omega<-1/3$ but we must be careful since with this
parameter we see that the temperature increases.

We can try to generalize this scenario taking into account various kinds of
matter. The idea is as follow. We can define a general energy density
$\widetilde{\rho}$ as:
\begin{equation}
\widetilde{\rho}=\sum_{i=0}^{6}\rho_{i}%
\end{equation}
where $\rho_{i}$ stands for each kind of energy density, and the parameter
$i=0,1,...,6$ in such a way that: $i=0$ correspond to $\omega=-1$ (the false
vacuum)$,$ $i=1$ correspond to domain walls i.e. to $\omega=-2/3$, $i=2$ to
$\omega=-1/3,$ $i=3$ to $\omega=0$, $i=4$ to $\omega=1/3,$ $i=5$ to
$\omega=2/3$ and finally $i=6$ to $\omega=1;$ but in such a way that each type
of matter verifies the relation
\begin{equation}
p_{i}=\omega_{i}\rho_{i}%
\end{equation}
in this way we define the total pressure as:
\begin{equation}
\widetilde{p}=\sum_{i=0}^{6}p_{i}%
\end{equation}
but in this case we do not impose that it is verified for each kind of matter
the relation:
\begin{equation}
\rho_{i}=A_{\omega_{i}}f^{-3\left(  \omega+1\right)  }%
\end{equation}

Our purpose is as follows: we impose that the relation $div(\widetilde{T})=0$
it is verified i.e.
\begin{equation}
\widetilde{\rho}^{\prime}+3(\widetilde{p}+\widetilde{\rho})H=0
\end{equation}
taken into account that $p_{i}=\omega_{i}\rho_{i}$ then
\begin{equation}
\widetilde{\rho}^{\prime}+3H\sum_{i=0}^{6}\left[  \left(  \omega
_{i}+1\right)  \rho_{i}\right]  =0
\end{equation}
that we can ``\emph{approximate}'' through the relation:
\begin{equation}
\widetilde{\rho}=A_{m}f^{-m}%
\end{equation}
with $m=3\sum_{i=0}^{6}\left(  \omega_{i}+1\right)  .$ It is proven in a
trivial way that if we consider only one type of matter we then recuperate
the above results i.e.. $m=3(\omega+1).$

Therefore with the next set of governing quantities $\frak{M}$=$\frak{M}$ $ (A_{m},B,t) $ we
arrive to obtain the following table of results:
\begin{equation}
\begin{array}
[c]{l}
G  \propto A_{m}^{\frac{2}{x+1}}B^{\frac{2}{x+1}}t^{\frac{4}{x+1}%
-2}\\
c   \propto A_{m}^{\frac{1}{x+1}}B^{\frac{1}{x+1}}t^{^{\frac{2}{x+1}-1}%
}\\
\Lambda   \propto A_{m}^{\frac{-2}{x+1}}B^{\frac{-2}{x+1}}t^{\frac{-4}{x+1}%
}\\
\hbar   \propto A_{m}^{\frac{3}{x+1}}B^{\frac{3}{x+1}-1}t^{\frac{6}{x+1}%
-1}\\
m_{i}   \propto A_{m}^{\frac{1}{x+1}}B^{\frac{-x}{x+1}}t^{\frac{2}{x+1}%
}\\
e^{2}\varepsilon_{0}^{-2}   \propto A_{m}^{\frac{4}{x+1}}B^{\frac{4}{x+1}%
-1}t^{\frac{8}{x+1}-2}\\
f   \propto A_{m}^{\frac{1}{x+1}}B^{\frac{1}{x+1}}t^{\frac{2}{x+1}%
}\\
\rho   \propto B^{-1}t^{-2}\\
k_{B}\theta  \propto A_{m}^{\frac{1}{x+1}}B^{\frac{3}{x+1}-1}t^{\frac
{6}{x+1}-2}\\
a^{-1/4}s   \propto A_{m}^{\frac{3}{x+1}}B^{\frac{3}{x+1}-\frac{3}{4}%
}t^{\frac{6}{x+1}-\frac{3}{2}}\\
q   =\frac{x+1}{2}-1
\end{array}
\end{equation}
where $x+1=m=3\sum_{i=0}^{6}\left(  \omega_{i}+1\right)  .$ If for example we
consider an universe with dust $\left(  \omega=0\right)  $ and radiation
$\left(  \omega=1/3\right)  $ then $m=9,$ we obtain a very surprising results as we can
see
\begin{equation}
G\propto t^{-8/5},\quad c\propto t^{-4/5},\quad\hbar\propto t^{-2/5}%
,\quad\Lambda\propto t^{-2/5},\quad\rho\propto t^{-2},\quad f\propto
t^{1/5}\quad etc...
\end{equation}

We have developed this section since when we attack in the next section
the Wheeler-DeWitt equation we shall see that in it we can take into account
all these kinds of matter, for this reason we need to know beforehand how 
all the constants in these frameworks vary.
.

\section{Wheeler-DeWitt equation. \textbf{Minisuperspace approximation.}}

\subsection{\textbf{Case }$\mathbf{1}$\textbf{. General case.}}

The Einstein equation has precisely the same form as the Hamiltonian for a
zero-energy particle whose position is described by a coordinate $f$. To
quantize we must replace the momentum conjugate to $f$ \ by its corresponding
\ operator, according to:
\begin{equation}
p_{f}\rightarrow\widehat{p}_{f}=-i\hbar\partial_{f}%
\end{equation}

The Einstein equations may be obtained via Hamilton's principle \cite{K}. From
the Einstein field equations we define our Lagrangian as:
\begin{equation}
L=-\aleph f^{3}\left[  \left(  \frac{f^{\prime}}{f}\right)  ^{2}-\frac{kc^{2}%
}{f^{2}}+\frac{8\pi G}{3c^{2}}(\widetilde{\rho})\right]
\end{equation}
where $\aleph=\frac{3\pi c^{2}}{4G}$ is a renormalization factor and
$\widetilde{\rho}$ stands for all the possible forms of energy density. In the
case of FRW universe the action takes the form:
\begin{equation}
S_{grav}=\int L_{grav}dt=\aleph\int f^{3}\left[  -\left(  \frac{f^{\prime}}%
{f}\right)  ^{2}+\frac{kc^{2}}{f^{2}}-\frac{8\pi G}{3c^{2}}(\widetilde{\rho
})\right]  dt
\end{equation}

The momentum conjugate to $f$ is:
\begin{equation}
p\equiv\frac{\partial L}{\partial f^{\prime}}=-2\aleph ff^{\prime}%
\end{equation}

We define the Hamiltonian $H(f^{\prime},f)$ as:
\begin{equation}
H=pf^{\prime}-L
\end{equation}
simplifying it yields
\begin{equation}
H=-\aleph f^{3}\left[  \left(  \frac{f^{\prime}}{f}\right)  ^{2}+\frac{kc^{2}%
}{f^{2}}-\frac{8\pi G}{3c^{2}}(\widetilde{\rho})\right]
\end{equation}
the Hamiltonian has been written in terms of $f^{\prime}$ to show explicitly
that it is identically zero and is not equal to the total energy $H=0$. In terms
of the conjugate momentum $p,$ the Einstein equations may be written as
follow:
\begin{equation}
H(p,f)=-\aleph f^{3}\left[  \frac{p^{2}}{4\aleph^{2}f^{4}}+\frac{kc^{2}}%
{f^{2}}-\frac{8\pi G}{3c^{2}}(\widetilde{\rho})\right]  =0
\end{equation}
which, of course is also equal to zero. Straightforward algebra it yields:
\begin{equation}
p^{2}+4\aleph^{2}f^{2}\left(  kc^{2}-\frac{8\pi G}{3c^{2}}f^{2}(\widetilde
{\rho})\right)  =0
\end{equation}
Quantizing, making the replacement
\begin{equation}
\widehat{p}\longrightarrow-i\hbar\partial_{f}%
\end{equation}
and imposing $\widehat{H}\psi=0$ results the Wheeler-DeWitt equation in the
minisuperspace approximation for arbitrary $k$ and with different kinds of
matter expressed in $\widetilde{\rho}.$

Now, if we take into account that $\aleph=\frac{3\pi c^{2}}{4G}$ then
\begin{equation}
\left[  \frac{d^{2}}{df^{2}}-\frac{9\pi^{2}c^{4}f^{2}}{4\hbar^{2}G^{2}}\left(
kc^{2}-\frac{8\pi G}{3c^{2}}(\widetilde{\rho})f^{2}\right)  \right]  \psi=0
\end{equation}
and if we replace $\widetilde{\rho}$ by (for example) $\widetilde{\rho}%
=\rho+\rho_{vac.}$ where $\rho$ stands for energy density of matter or
radiation and $\rho_{vac.}$ is the energy density of the vacuum expressed in
terms of the cosmological constant
\begin{equation}
\rho_{vac.}=\frac{\Lambda c^{4}}{8\pi G}%
\end{equation}
Note that other types of matter can be taken into account like for example
strings, ultrastiff matter, domains walls etc... \ algebra brings us to the
following expression
\begin{equation}
\left[  \frac{d^{2}}{df^{2}}-\frac{9\pi^{2}c^{6}f^{2}}{4\hbar^{2}G^{2}}\left(
k-\frac{\Lambda f^{2}}{3}-\frac{8\pi G}{3c^{4}}A_{\omega}f^{-1-3\omega
}\right)  \right]  \psi=0\label{eq14}%
\end{equation}
where we have used the relation between $\rho$ and $f$
\begin{equation}
\rho=A_{\omega}f^{-3(\omega+1)}%
\end{equation}
or in a compact notation
\begin{equation}
\frac{d^{2}\psi}{df^{2}}-V(f)\psi=0
\end{equation}
where
\begin{equation}
V(f)=\frac{9\pi^{2}f^{2}}{4l_{p}^{4}}\left(  k-\frac{\Lambda f^{2}}{3}%
-\frac{8\pi G}{3c^{4}}A_{\omega}f^{-1-3\omega}\right)
\end{equation}
being $l_{p}$ Planck's length.

The W-DW eq. is identical to one-dimensional time-independent Schr\"{o}dinger
eq. for a one-half unit mass particle of zero energy subject to the potential
\begin{equation}
V(f)=\frac{9\pi^{2}}{4l_{p}^{4}}f^{2}\left(  k-\frac{\Lambda f^{2}}{3}%
-\frac{8\pi G}{3c^{4}}A_{\omega}f^{-1-3\omega}\right)
\end{equation}

The ``particle'' at $f=0$ - a quantum FRW universe of zero radius or indeed
our cosmological ``nothing''- may quantum mechanically tunnel through the
potential barrier to appear at $f=f_{0}$. This tunneling event represents a FRW
universe of size (scale factor) $f_{0}$ that has quantum mechanically popped
into existence i.e. a universe that has been created spontaneously and
nonsingularly (since $f_{0}\neq0).$ Choosing the tunneling wave function, it
is now a simple matter to calculate the probability with which this occurs. If
we denote the amplitude for the quantum creation by
\begin{equation}
\left|  <FRW(f_{0})|nothing>\right|  ^{2}=P
\end{equation}%
\begin{equation}
P\backsimeq\exp\left[  -\frac{2}{\hbar}\int_{0}^{f_{0}}\sqrt{2m_{p}\left[
V(f^{\prime})\right]  }df^{\prime}\right]
\end{equation}
The tunneling probability of the universe will be obtained from the expression
above (see \cite{ATK},\cite{Fil}).

\subsection{\textbf{The dimensional stone.}}

We now calculate the multiplicity of the dimensional base\textbf{\ }\cite{C}.
For this purpose we observe that equation (\ref{eq14})
\begin{equation}
\frac{\partial^{2}\psi}{\partial f^{2}}-\frac{9\pi^{2}c^{4}f^{2}}{4\hbar
^{2}G^{2}}\left(  kc^{2}-\frac{\Lambda c^{2}f^{2}}{3}-\frac{8\pi G}{3c^{2}%
}A_{\omega}f^{-1-3\omega}\right)  \psi=0
\end{equation}
can be written in the following dimensionless products as:
\begin{equation}
\pi_{1}:=\frac{c^{6}f^{4}}{\hbar^{2}G^{2}}\qquad\pi_{2}:=\frac{c^{6}%
f^{6}\Lambda}{\hbar^{2}G^{2}}\qquad\pi_{3}:=\frac{c^{2}A_{\omega}%
f^{3(1-\omega)}}{\hbar^{2}G}\label{Pi1}%
\end{equation}
from $\pi_{1}$ it is observed that $l_{p}\propto f,$ from $\pi_{2}$ and
$\pi_{1}$ we can see that $\Lambda\propto f^{-2}$ and from $\pi_{3}$ it is
observed that $\hbar\propto A_{\omega}^{1/2}B^{-1/2}f^{3(1-\omega)/2}$ . We
shall see in section $4$ that these results are precisely what we obtain when
the possibility of time-varying constants in GR is considered in section $2$
but with $\beta=0$. We would like to emphasize that while in these cases it
was studied the variation of the ``constants'' in semiclassical cosmological
models where constant $\hbar$ was introduced through an equation of state and
to extrapolate the result to any kind of matter the result obtained here does
not depend on any equation of state. It is observed that it only depends on
the kind of matter, this fact is reflected in constant $A_{\omega}$. In this
way we validate our extrapolation as well as the obtained results in section
$2$. With these relations (\ref{Pi1}) we can say that the Wheeler-DeWitt
equation remains gauge invariant, see (\cite{T1})

We proceed to calculate the multiplicity of the base in this model. The rank
of the matrix of the exponents of the quantities and constants included in the
monomia is $3$ as it results immediately from:
\begin{equation}%
\begin{array}
[c]{l|rrrrcc}%
& G & c & \hbar & \Lambda &  A_{\omega} & f\\\hline
\pi_{1} & -2 & 6 & -2 & 0 & 0 & 4\\
\pi_{2} & -2 & 6 & -2 & 1 & 0 & 6\\
\pi_{3} & -1 & 2 & -2 & 0 & 1 & 3(1-\omega)
\end{array}
\end{equation}
The multiplicity of the dimensional base is therefore $m=($\emph{number of
quantities and constants)-(rank of the matrix) } is $3$. Thus we can use as
base $B=\left\{  G,c,\hbar\right\}  $. The only fundamental quantity is $f$
and the set of unavoidable constants are $\mathcal{C}$$=\left\{
\Lambda,A_{\omega},G,c,\hbar\right\}  .$ It is observed that the set of
fundamental constants $\left[  G,c,\hbar\right]  $ form the famous Planck's
system of units and this possibility should be taken into account.

Once the dimensional base of the theory (model) is obtained we go next to
calculate the dimensional equation of each quantity, these are:%

\begin{equation}
\left[  G\right]  =G\qquad\left[  c\right]  =c\qquad\left[  \hbar\right]
=\hbar\qquad\left[  k_{B}\right]  =k_{B}%
\end{equation}
since we can take into account the equation of state $\rho=a\theta^{4}$ where
$a$ stands for the radiation constant ($a\propto k_{B}^{4}/\hbar^{3}c^{3}).$
The dimensional equations of the rest of the quantities are:
\begin{equation}%
\begin{array}
[c]{lll}%
\left[  f\right]  =\hbar^{1/2}c^{-3/2}G^{1/2} &  & \left[  \rho\right]
=\hbar^{-1}c^{7}G^{-2}\\
\left[  A_{0}\right]  =\hbar^{1/2}c^{1/2}G^{-1/2} &  & \left[  t\right]
=\hbar^{1/2}c^{-5/2}G^{1/2}\\
\left[  \theta\right]  =\hbar^{1/2}c^{5/2}G^{-1/2}k_{B}^{-1} &  & \left[
V(f)\right]  =\hbar^{-1}c^{3}G^{-1}%
\end{array}
\end{equation}%
\begin{equation}
\left[  \psi\right]  =\hbar^{-n/4}c^{3n/4}G^{-n/4}\qquad\qquad from \qquad
\int\psi\psi^{\ast}\ast d\Omega=1
\end{equation}
where $\ast d\Omega$ is the volume-element on superspace, $\ast$ being the
Hodge dual in the supermetric. In this case it is:
\begin{equation}
\left[  \psi\right]  =\hbar^{-1/4}c^{3/4}G^{-1/4}%
\end{equation}
for more details about the dimensions of the wave function see (\cite{AR})

It is trivially observed that:
\begin{equation}%
\begin{array}
[c]{lll}%
\left[  f\right]  =l_{p} &  & \left[  \rho\right]  =\rho_{p}\\
\left[  A_{0}\right]  =m_{p} &  & \left[  t\right]  =t_{p}\\
\left[  \theta\right]  =\theta_{p} &  & etc...
\end{array}
\end{equation}
i.e. the dimensional equation of $f$ i.e. $\left[  f\right]  $ is precisely
the definition of the length of Planck etc... \ and which is important, we
obtain the equivalence between $B=\left\{  G,c,\hbar,k_{B}\right\}  $ and
$B=\left\{  L,M,T,\theta\right\}  $ i.e. between the obtained base from the
Wheeler-DeWitt equation (derived from Einstein's equations) and the obtained
one from the traditional Friedmann's equations. It is observed that the Planck
system of units acquires full sense within this framework.

\subsection{\textbf{Schr\"{o}dinger reduction.}}

Note that the Wheeler-DeWitt equation is defined on curved ``space-time'' but
it is possible to reduce it to the Schr\"{o}dinger one in an effective flat
space which permits the conventional quantum-mechanical procedure to be used
(for example the calculation of the penetration factor). Therefore the
Schr\"{o}dinger equation after straightforward algebra yields\cite{Fil}:
\begin{equation}
\frac{\hbar^{2}}{2m_{p}}\frac{d^{2}\psi}{df^{2}}-\frac{9\pi^{2}}{8}m_{p}%
c^{2}\frac{f^{2}}{l_{p}^{2}}\left(  k-\frac{\Lambda f^{2}}{3}-\frac{8\pi
BA_{\omega}}{3c^{2}}f^{-1-3\omega}\right)  \psi=0,\label{SCH}%
\end{equation}
where $B$ is $B=G/c^{2}$ and the potential, in this case, is defined by:
\begin{equation}
U(f)=\frac{9\pi^{2}}{8}m_{p}c^{2}\frac{f^{2}}{l_{p}^{2}}\left(  k-\frac
{\Lambda f^{2}}{3}-\frac{8\pi BA_{\omega}}{3c^{2}}f^{-1-3\omega}\right)  ,
\end{equation}
it is observed that now this potential has dimensions of energy. Writing it in
a dimensionless form we obtain the following $\pi-monomia.$%
\begin{equation}
\tilde{\pi}_{1} =\frac{m_{p}^{2}c^{2}f^{4}}{\hbar^{2}l_{p}^{2}}\qquad
\tilde{\pi}_{2} =\frac{m_{p}^{2}c^{2}f^{6}\Lambda}{\hbar^{2}l_{p}^{2}}%
\qquad\tilde{\pi}_{3} =\frac{m_{p}^{2}BA_{\omega}f^{3(1-\omega)}}{\hbar
^{2}l_{p}^{2}}%
\end{equation}
simplifying them it is observed that they are the same monomia that the
obtained ones in equation (\ref{Pi1}). Therefore this new equation has the
same dimensional base that the obtained one in the case of the Wheeler-DeWitt
equation. We shall use equation (\ref{SCH}) in all our calculations.

\subsection{Case $2$. \textbf{Scalar Field Model.}}

The density and pressure for an interacting scalar field $\phi$ is:
\begin{equation}
\rho_{\phi}=\frac{1}{2}{\dot{\phi}}^{2}+V(\phi)\label{3-1}%
\end{equation}
and
\begin{equation}
p_{\phi}=\frac{1}{2}{\dot{\phi}}^{2}-V(\phi)\label{3-2}%
\end{equation}
where $V(\phi)$ is the interaction potential. Assuming that the scalar field
dominates, the field equation becomes
\begin{equation}
H^{2}\equiv\left(  \frac{f^{\prime}}{f}\right)  ^{2}=\frac{8\pi G}{3c^{2}%
}\left[  \frac{1}{2}{\dot{\phi}}^{2}+V(\phi)\right] \label{3-3}%
\end{equation}
and instead of \ the ordinary conservation principle, one obtains (assuming a
massless field and ignoring spatial derivatives)
\begin{equation}
\ddot{\phi}+3H\dot{\phi}+V^{\prime}=0\label{3-4}%
\end{equation}
which is the equation of motion for the scalar field where $V^{\prime}%
\equiv\frac{dV}{d\phi}$.

The usual way of quantizing the FRW model for a scalar field is to treat
$\phi$ and $f$ as independent variables each with their own canonical momenta.

We define our Lagrangian as:
\begin{equation}
L=2\pi^{2}\left\{  \left(  -\frac{3c^{2}}{8\pi G}\right)  \left[  f\left(
f^{\prime}\right)  ^{2}-fc^{2}K\right]  +\frac{1}{2}f^{3}\dot{\phi}^{2}%
-f^{3}V(\phi)\right\}
\end{equation}
from the field equation:
\begin{equation}
\left(  \frac{f^{\prime}}{f}\right)  ^{2}=\frac{8\pi G}{3c^{2}}\left[
\frac{1}{2}{\dot{\phi}}^{2}+V(\phi)\right]
\end{equation}
the standard procedure brings us to the following expressions:
\begin{equation}
\frac{\partial L}{\partial f^{\prime}}=\frac{-3\pi}{2}\frac{c^{2}ff^{\prime}%
}{G}:=\pi_{f}%
\end{equation}%
\begin{equation}
\frac{\partial L}{\partial{\dot{\phi}}}=2\pi^{2}f^{3}\dot{\phi}:=\pi_{\phi}%
\end{equation}
we define the Hamiltonian as:
\begin{equation}
H=\pi_{f}f^{\prime}+\pi_{\phi}\dot{\phi}-L
\end{equation}
after some simplifications straightforward algebra it yields:
\begin{equation}
H=-\frac{G\pi_{f}^{2}}{3\pi c^{2}f}+\frac{\pi_{\phi}^{2}}{4\pi f^{3}}+2\pi
^{2}f^{3}V(\phi)-\frac{3\pi c^{4}fK}{4G}%
\end{equation}%
\begin{equation}
H=-\pi_{f}^{2}+\frac{3c^{2}}{4\pi G}\frac{\pi_{\phi}^{2}}{f^{2}}+\frac{6\pi
c^{2}}{G}f^{4}V(\phi)-\frac{9\pi^{2}c^{6}f^{2}K}{4G^{2}}%
\end{equation}

Canonical quantization it yields if:
\begin{equation}
\widehat{\pi}_{f}\longmapsto-i\hbar\partial_{f}%
\end{equation}%
\begin{equation}
\widehat{\pi}_{\phi}\longmapsto-i\hbar\partial_{\phi}%
\end{equation}
therefore the Wheeler-DeWitt equation in this case becomes:
\begin{equation}
\left[  -\frac{{\partial}^{2}}{\partial f^{2}}+\frac{3c^{2}}{4\pi G}\frac
{1}{f^{2}}\frac{{\partial}^{2}}{\partial{\phi}^{2}}+U(f,\phi)\right]
\Psi=0\label{3-8}%
\end{equation}
where the potential is
\begin{equation}
U(f,\phi)=\frac{9{\pi}^{2}c^{6}}{4\hbar^{2}G^{2}}f^{2}\left[  k-\frac{8\pi
G}{3c^{4}}V(\phi)f^{2}\right] \label{3-9}%
\end{equation}

\subsection{\textbf{The dimensional stone.}}

We now calculate the multiplicity of the dimensional base. For this purpose we
observe that equation (\ref{3-8}) can be written in the following
dimensionless products as:
\begin{equation}
\pi_{1} :=\frac{c^{2}}{G{\phi}^{2}}\qquad\pi_{2} :=\frac{c^{6}f^{4}}{\hbar
^{2}G^{2}}\qquad\pi_{3} :=\frac{c^{2}V(\phi)f^{6}}{\hbar^{2}G}%
\end{equation}
We proceed to calculate the multiplicity of the base in this model. The rank
of the matrix of the exponents of the quantities and constants included in the
monomia is $3$ as it results immediately from:
\begin{equation}%
\begin{array}
[c]{l|rrrrrr}%
& G & c & \hbar & \phi &  V & f\\\hline
\pi_{1} & -1 & 2 & 0 & -2 & 0 & 0\\
\pi_{2} & -2 & 6 & -2 & 0 & 0 & 4\\
\pi_{3} & -1 & 2 & -2 & 0 & 1 & 6
\end{array}
\end{equation}
The multiplicity of the dimensional base is therefore $m=($\emph{number of
quantities and constants)-(rank of the matrix) }in this case is $3$. Thus we
can use as base, for example $B=\left\{  G,c,\hbar\right\}$. The fundamental
quantities are $\left[  f,\phi\right]  $ and the set of unavoidable constants
are $\mathcal{C}$$=\left\{  G,c,\hbar\right\}$, it is observed that the set
of fundamental constants $\left[  G,c,\hbar\right]  $ form the famous Planck's
system of units.

The dimensional equations of the rest of the quantities are:
\begin{equation}
\left[  \phi\right]  =cG^{-1/2}\qquad\qquad\left[  V\right]  =\hbar^{-1}%
c^{7}G^{-2}%
\end{equation}

\section{Variable constants.}

Following the results obtained in section $2$ (with $\beta=0)$ (see \cite{T1})
we determine the behaviour of the constants in function on $f$ since this
quantity is the only fundamental quantity in the theory of quantum cosmology
(in our approximation in this minisuperspace). As we mentioned previously, this
behaviour depends on the equation of state. A simple algebra exercise brings
us to obtain the next solutions where now the set of governing parameters is:
$\frak{M=M}$$\left\{  A_{\omega},B,f\right\}  $
\begin{equation}%
\begin{array}
[c]{l}%
G\propto A_{\omega}B^{2}f^{1-x}\\
c\propto A_{\omega}^{1/2}B^{1/2}f^{(1-x)/2}\\
\hbar\propto A_{\omega}^{1/2}B^{-1/2}f^{(5-x)/2}\\
m_{i}\propto B^{-1}f\\
\Lambda\propto f^{-2}%
\end{array}
\label{R1}%
\end{equation}
where $x=3\omega+2$ in the usual notation. It is interesting to emphasize the
next relations obtained from eq. (\ref{R1}):
\begin{equation}
l_{p}\thickapprox f
\end{equation}
and that $\Lambda\propto f^{-2}$ $\left(  \Lambda=d_{0}f^{-2},d_{0}%
\in\mathbb{R}\right)  $ and $m_{i}\propto f$ in all cases i.e. these results
do not depend on the equation of state. As we mentioned earlier all these
relationships are the same as the ones obtained in equation (\ref{Pi1})

For example, if we impose $\omega=1/3$ it is obtained:
\begin{equation}%
\begin{array}
[c]{l}%
G \propto A_{\omega}B^{2}f^{-2}\\
c \propto A_{\omega}^{1/2}B^{1/2}f^{-1}\\
\hbar\propto A_{\omega}^{1/2}B^{-1/2}f
\end{array}
\label{R2}%
\end{equation}
$\Lambda\propto f^{-2}$ and $m_{i}\propto f,$ and if we impose $\omega=-1/3 $
in this case we obtain:
\begin{equation}%
\begin{array}
[c]{l}%
G =const.\\
c =const.\\
\hbar\propto A_{\omega}^{1/2}B^{-1/2}f^{2}%
\end{array}
\label{R3}%
\end{equation}
while $\Lambda\propto f^{-2}$ and $m_{i}\propto f.$

If the constants $G,c,\hbar$ and $\Lambda$ vary, then the Wheeler-DeWitt
equation
\[
\frac{d^{2}\psi}{df^{2}}-\frac{9\pi^{2}f^{2}}{4l_{p}^{4}}\left(
k-\frac{\Lambda f^{2}}{3}-\frac{8\pi G}{3c^{4}}A_{\omega}f^{-1-3\omega
}\right)  \psi=0
\]
it yields:
\begin{equation}
\left[  \frac{d^{2}}{df^{2}}-\frac{9\pi^{2}}{4f^{2}}\left(  k-\frac{d_{0}}%
{3}-\frac{8\pi d_{\omega}}{3}\right)  \right]  \psi=0\label{sil1}%
\end{equation}
where the potential is defined by:
\begin{equation}
V(f)=\frac{9\pi^{2}}{4f^{2}}\left(  k-\frac{d_{0}}{3}-\frac{8\pi d_{\omega}%
}{3}\right) \label{pot1}%
\end{equation}
and where $d_{\omega}\in\mathbb{R}$ since
\begin{equation}%
\begin{array}
[c]{l}%
l_{p} \propto f\\
\Lambda\propto f^{-2}\\
G/c^{2} =B\\
c^{2} \propto A_{\omega}Bf^{-1-3\omega}%
\end{array}
\end{equation}
while the Schr\"{o}dinger equation it yields:
\begin{equation}
A_{\omega}f^{2-3\omega}\frac{d^{2}\psi}{df^{2}}-\frac{9\pi^{2}}{4}A_{\omega
}f^{-3\omega}\left(  k-\frac{d_{0}}{3}-\frac{8\pi d_{\omega}}{3}\right)
\psi=0\label{SCHV}%
\end{equation}
where the potential is defined by:
\begin{equation}
U(f)=\frac{9\pi^{2}}{4}A_{\omega}f^{-3\omega}\left(  k-\frac{d_{0}}{3}%
-\frac{8\pi d_{\omega}}{3}\right) \label{pot2}%
\end{equation}
if we simplify eq. (\ref{SCHV}) then it reduces to:
\begin{equation}
\frac{d^{2}\psi}{df^{2}}-\frac{9\pi^{2}}{4f^{2}}\left(  k-\frac{d_{0}}%
{3}-\frac{8\pi d_{\omega}}{3}\right)  \psi=0\label{sil2}%
\end{equation}
it is observed that in this case both equations (\ref{sil1}) and (\ref{sil2})
are identical.

For example in the case of $\omega=1/3$ equation (\ref{SCHV}) it yields:
\begin{equation}
A_{\omega}f\frac{d^{2}\psi}{df^{2}}-\frac{9\pi^{2}}{4}A_{\omega}f^{-1}\left(
k-\frac{d_{0}}{3}-\frac{8\pi d_{\omega}}{3}\right)  \psi=0
\end{equation}
and if we use $\omega=-1/3$ it yields:
\begin{equation}
A_{\omega}f^{3}\frac{d^{2}\psi}{df^{2}}-\frac{9\pi^{2}}{4}A_{\omega}f\left(
k-\frac{d_{0}}{3}-\frac{8\pi d_{\omega}}{3}\right)  \psi=0
\end{equation}
Note that the value of the numerical constants $d_{i}\in\mathbb{R}$ are fundamental.

We would like to emphasize that in this paper all the constants are considered
as variable, including the Planck's constant while in previous references it
is considered only the variation of the constant $\Lambda$ or the variation of
$G,c$ and $\Lambda$ but no the Planck's constant $\hbar$ \cite{Nor,HAR}

\subsection{\textbf{Quantum tunneling.}}

The tunneling probability follows from the expression:
\begin{equation}
P\backsimeq\exp\left[  -\frac{2}{\hbar}\int_{0}^{f_{0}}\sqrt{2m_{p}%
U(f^{\prime})}df^{\prime}\right]
\end{equation}
where $U(f^{\prime})$ follows from expression (\ref{pot2}), since until now all the
proposals only consider the potential $V(f)$ (\ref{pot1}) despising
all the ``constants''. This expression is obviously dimensionally homogeneous
while the usual ones are not.

Our potential is defined by:
\begin{equation}
U(f)=\frac{9\pi^{2}}{4}A_{\omega}f^{-3\omega}\left(  k-\frac{d_{0}}{3}%
-\frac{8\pi d_{\omega}}{3}\right)
\end{equation}
as we have seen above the ``constants'' $m_{p}$ and $\hbar$ vary as:
\begin{equation}
m_{p}\propto B^{-1}f\qquad\qquad\qquad\hbar\propto A_{\omega}^{1/2}%
B^{-1/2}f^{3(1-\omega)/2}%
\end{equation}
therefore this probability may approximate by:%

\begin{equation}
P\propto\exp\left[  -\frac{2\sqrt{\tilde{\chi}}}{3-3\omega}\right]
\end{equation}
where $\tilde{\chi}=6\pi^{2}\left(  3k-d_{0}-8\pi d_{\omega}\right)  $, we see
that such probability only depends on the considered equation of state. If for
example we take $\omega=1$ (ultrastiff matter) then the probability vanished.
This formula may help us to find an equation of state in a pre-big-bang scenario.

\section{Conclusions}

In this work we have try to show how all the constants vary when the
conservation principle is taken into account. We have started revising previous
results and then we have gone next to incorporate in our equations the condition
$divT=0.$ This condition allows us to take into account mechanics of adiabatic
matter creation in such a way that we have been able to get rid of the entropy
problem. We have show how all the ``constants'' vary as well as all the
physical quantities in the model in function of the equation of state and
trying to generalize our results we have contemplated the possibility of
considering a multi-fluid frameworks since these possibilities are present in
the Wheeler-DeWitt equation. To study the variation of ``constant'' $\hbar,$
we need to introduce it into the equations by imposing a special equation of
state and to make assumptions about its behaviour but with the hypothesis of
the conservation principle i.e. taking into account the condition $divT=0,$ we
have seen that no previous hypotheses are needed. To validate all these results
we have considerated the W-DW equation.

We have studied from a dimensional point of view the Wheeler-DeWitt equation
in the minisuperspace approximation. We have shown that the set of governing
parameters is formed by $\wp=\left(  f,G,c,\hbar,\Lambda,A_{\omega}\right)  $
(depending in each case) where $f$ is the only fundamental quantity and the
set of unavoidable constants $\mathcal{C}$ is constituted by $\mathcal{C}%
$$=\left\{  G,c,\hbar,\Lambda,A_{\omega}\right\}.$ A remarkable and
surprising feature of the theory is the fact that it is independent of time, the only fundamental quantity is the scale factor $f.$ As we have shown the
multiplicity of the dimensional base is usually $3$ and that a possible base
could be formed by the Planck's system of units i.e. $B=\left\{
G,c,\hbar,k_{B}\right\}  $ showing in this way that is precisely in this
framework where this system acquires a complete sense. Writing the
Wheeler-DeWitt equation in a dimensionless way we have seen that the behaviour
of the constants is the same that the obtained ones in the case of the
standard cosmology. In this way we have deduced the Wheeler-DeWitt equation
with all the constants varying as well as the Schr\"{o}dinger equation (its
reduced form). We have also seen that our model may arise via quantum tunneling.

\bigskip

\textbf{{\LARGE A}cknowledgement} I wish to express my gratefulness to Javier
Aceves for his collaboration in the translation into English.\newline

\nonumsection{References}

\end{document}